\documentclass{article}
\usepackage[utf8]{inputenc}
\usepackage{amssymb, amsmath, amsfonts, xcolor, verbatim, listings, fullpage, physics, mathrsfs, subcaption}
\usepackage{graphicx, hyperref, bbm}%
\usepackage[backend=biber, style=apa ,natbib=true, sortcites]{biblatex}
\newcommand{\indep}{\perp \! \! \! \perp}

\usepackage[linesnumbered , ruled, noend, norelsize]{algorithm2e}
  \SetKwInOut{Input}{Input}
  \SetKwInOut{Output}{Output}
\IncMargin{1em}

\newcommand{\calT}{\mathcal{T}}
\newcommand{\calC}{\mathcal{C}}
\newcommand{\calS}{\mathcal{S}}

\newtheorem{theorem}{Theorem}
\newtheorem{definition}[theorem]{Definition}
\newtheorem{example}[theorem]{Example}

\title{Beyond Conjugacy for Chain Event Graph Model Selection}
\author{Aditi Shenvi and Silvia Liverani}
\date{\today}

\begin{document}

\maketitle

\begin{abstract}
    Chain event graphs are a family of probabilistic graphical models that generalise Bayesian networks and have been successfully applied to a wide range of domains. Unlike Bayesian networks, these models can encode context-specific conditional independencies as well as asymmetric developments within the evolution of a process. More recently, new model classes belonging to the chain event graph family have been developed for modelling time-to-event data to study the temporal dynamics of a process. However, existing model selection algorithms for chain event graphs and its variants rely on all parameters having conjugate priors. This is unrealistic for many real-world applications. In this paper, we propose a mixture modelling approach to model selection in chain event graphs that does not rely on conjugacy. Moreover, we also show that this methodology is more amenable to being robustly scaled than the existing model selection algorithms used for this family. We demonstrate our techniques on simulated datasets.   
\end{abstract}

\section{Introduction}
\label{sec:introduction}

Chain event graphs (CEGs) are a family of probabilistic graphical models that were first proposed in \citet{smith2008conditional} as an alternative to the family of Bayesian networks (BNs). In particular, CEGs were developed to explicitly accommodate processes exhibiting asymmetries of two types: (1) asymmetric independence structures or \textit{context-specific} conditional independences where some statistical independences hold for certain values of the conditioning variables but not the others; and (2) asymmetric event spaces which are precisely event spaces that do not admit a product space structure. The latter asymmetry arises due to the presence of structural zeros and structural missing values, often-times by design \citep{shenvi2020constructing}. For example, consider modelling hospitalisations arising from infection caused by a circulating virus, and suppose that one of the two strains (call it strain A) of the virus has no treatment currently available while the other has a choice of two possible treatments. On the one hand, a variable of ``Treatment" with state space $\{\text{Treatment 1}, \text{Treatment 2}\}$ would be structurally missing and have no sensible value for those infected by strain A of the virus. Whereas on the other hand, if its state space is redefined to be $\{\text{Treatment 1}, \text{Treatment 2}, \text{No treatment}\}$ then $\text{Treatment 1}$ and $\text{Treatment 2}$ would have structurally zero counts for those infected by strain A, i.e. irrespective of the sample size, there would always be zero individuals who are treated with either  $\text{Treatment 1}$ or $\text{Treatment 2}$ among those infected by strain A. Such a process is inherently asymmetric. BNs, being variable-based -- i.e. they use variables as the building blocks of their models -- are unable to fully describe such asymmetries within their underlying statistical model and graphical structure. The CEG for a process, on the other hand, is obtained through a transformation of an event tree describing the process and thus, has an event-based\footnote{An event is an element or a subset of elements of the state space of a variable.} topology. This event-based formulation enables CEGs to fully embed structural asymmetries within its model and graph. In fact, in order to accommodate such asymmetries within a BN model, the modifications proposed in the literature typically rely on tree-based structures, see e.g. \citet{boutilier1996context, zhang1999role, poole2003exploiting, jabbari2018instance}.

The parameters of a \textit{vanilla CEG}\footnote{In the CEG literature, a `CEG' often refers to the simplest discrete state space class of this family. Here, we refer to this as a `vanilla CEG' to distinguish it from other classes of the CEG family.} are given by the parameters of the conditional transition distributions for the nodes in its graph. These distributions govern which event occurs next given that a particular node has been reached. More recently, new classes of the CEG family have been proposed for modelling time-to-event data to study the temporal dynamics of a process \citep{shenvi2021non, shenvi2019bayesian, barclay2015dynamic, collazo2018n}. For brevity, we shall call these CEGs as \textit{temporal CEGs}. Temporal CEGs introduce conditional holding time random variables for the transitions modelled by the process. These random variables describe how long it takes for the next event to occur given that a particular node has been reached. For instance, if we consider the infection example introduced earlier, a temporal CEG would be suitable if we are not only interested in studying the evolution of the infection-to-hospitalisation trajectory of individuals but also how long it takes for the various transitions to occur in each trajectory. Thus, a model belonging to one of these classes has precisely two categories of parameters; one for modelling the conditional transition distributions, and the other for modelling the conditional holding time distributions. Note that the model selection exercise in these CEGs can be done independently for each category of random variable under the standard assumption of parameter independence. 

The existing Bayesian model selection algorithms employed for vanilla and temporal CEGs \citep{silander2013dynamic, cowell2014causal, freeman2011bayesian, shenvi2021non, strong2022bayesian} rely on the parameters of the conditional transition and conditional holding time distributions having conjugate priors. For instance, the Binomial or Multinomial distributions are typically used for the conditional transition distributions whereas the Weibull distribution with known shape parameter is used for the conditional holding times. 

Whilst conjugacy of prior and posterior distributions of parameters is desirable for its closed form analytical solutions and for the interpretability it lends to the hyperparameters, conjugate settings are either infeasible or inappropriate in most cases. Under the setting of sampling with replacement from a given population size with a fixed and finite number of categories, the Multinomial distribution (or equivalently, the Binomial distribution when the number of categories is 2) is perhaps the most appropriate choice for the conditional transition distributions \citep{minka2003bayesian}. However, there is no reason why the conditional holding time distributions need to belong to the conjugate family. A simple example here is that even if we believe the conditional holding times to be governed by a Weibull distribution, it is typically unlikely that we know the shape parameter of this distribution. Thus, the conjugacy requirement is less restrictive for vanilla CEGs than for temporal CEGs.

Moreover, even without consideration of the conjugacy issue, the existing model selection approaches are not easily scalable or are not robust when scaled. The two main existing approaches to model selection in vanilla and temporal CEGs are the brute-force approach of finding the globally optimal model and the agglomerative hierarchical clustering (AHC). The former approach is clearly not scalable, whereas the AHC is a greedy algorithm that can be scaled relatively well but it is not robust. 

In this paper, we propose a novel methodology for model selection in CEGs which casts the model selection problem into the problem of fitting a mixture model. We demonstrate that this simple change of perspective on the problem allows us to use well-developed and well-tested existing software such as Stan to support model selection in temporal CEGs with non-conjugate conditional holding time distributions. Further, we demonstrate how this approach enables a more robust scaling of model selection for vanilla and temporal CEGs, compared to the existing model selection algorithms, for conditional transition distributions when these follow the Binomial distribution. Thus, our paper vastly extends the range of applications that can be supported by the CEG family and also opens new avenues to extend their applicability. 

This paper is organised as follows. In Section \ref{sec:preliminaries} we review vanilla and temporal CEGs, and the model selection algorithms employed for these within the literature. In Section \ref{sec:model_selection} we describe how the model selection problem can be posed as a mixture modelling problem and discuss its advantages. In Section \ref{sec:experiments} we illustrate this methodology through simulated examples. We conclude with a discussion in Section \ref{sec:discussion}.

\section{Preliminaries} \label{sec:preliminaries}

\subsection{Chain Event Graphs}
\label{subsec:ceg}

CEGs are an event-based probabilistic graphical modelling family that describe the evolution of a process through a sequential unfolding of events. They harness the symmetries within the process to provide a compact representation of the process. Crucially, through their event-based formulation, they are able to embed asymmetric independence structures and asymmetric event spaces within their statistical models and graphs; see \citet{collazo2018chain, shenvi2018modelling, shenvi2021non}. 

The construction of a CEG model begins by eliciting an event tree description of the process from a combination of domain experts, existing literature and data. Event trees provide a natural framework for describing the step-by-step evolution of a process -- an excellent exposition of trees and their fundamental role in probability theory and causality can be found in \citet{shafer1996art}. A non-technical summary \citep{shenvi2020constructing} of the transitions an event tree must go through to become the graph of a CEG model are given below:

\begin{itemize}
    \itemsep0em
    \item Nodes in the event tree whose one-step-ahead evolutions are equivalent -- in terms of the conditional transition distributions for vanilla CEGs and the conditional transition and conditional holding time distributions for the temporal CEGs -- are said to be in the same \textit{stage} and are assigned the same colour to indicate their shared stage membership. 
    \item Nodes whose rooted subtrees (i.e. the subtree obtained by considering that node as the root) are isomorphic, in the structure and colour preserving sense, are said to be in the same \textit{position} and are merged into a single node which retains the colouring of its merged nodes.
    \item All the leaves of the tree are merged into a single node called the \textit{sink} node.
\end{itemize}

The simplest CEG class \citep{collazo2018chain}, which we refer to as the vanilla CEG here, explicitly models the conditional transition distributions but not the conditional holding time distributions -- these are included implicitly through its Markov assumption \citep{shenvi2021non}. Newer CEG classes such as the dynamic CEG \citep{barclay2015dynamic, collazo2018n}, extended dynamic CEG \citep{barclay2015dynamic} and the continuous-time dynamic CEG \citep{shenvi2019bayesian, shenvi2021non} were proposed for modelling longitudinal temporal processes with asymmetries. These classes explicitly model the conditional holding time distributions. We note here that these model classes can also be defined over non-longitudinal temporal processes -- which we define here to be a temporal process whose underlying event tree description is finite. For simplicity of illustration, in this paper we focus on these non-longitudinal temporal CEGs, referred to simply as temporal CEGs. The model selection approach described in this paper extends to dynamic temporal CEGs in a straightforward way. Further, under the standard assumption of parameter independence described in Section \ref{subsec:likelihood_separation}, the model selection exercise simplifies into two independent clustering problems; one for the conditional transition distributions and the other for the conditional holding time distributions. Therefore, we will focus on temporal CEGs with the understanding that our model selection approach for conditional transition distributions can be applied directly to vanilla CEGs as well.

Denote by $\calT$ an event tree with a finite node set $V(\calT)$ and a directed edge set $E(\calT)$. Each edge $e \in E(\calT)$ is an ordered triple of the type $(v, v', l)$ denoting that $e$ emanates from node $v$, terminates in node $v'$ and has edge label $l$. The set of leaves in $\calT$ is denoted by $L(\calT)$, and the non-leaf nodes known as \textit{situations} are represented by the set $S(\calT) = V(\calT) \backslash L(\calT)$. The set of children of a node $v$ is denoted by $\textmd{ch}(v)$. Let $\pmb{\Phi}_\calT = \{\pmb{\theta}_v |v \in S(\calT)\}$ where $\pmb{\theta_v} = (\theta(e) | e = (v, v', l) \in E(\calT), v' \in \textmd{ch(v)})$ denotes the conditional transition parameters for each node $v \in S(\calT)$. 

Each transition from node $v$ to $v'$ along some edge $e = (v, v', l)$ between them is associated with a holding time which indicates the time spent in node $v$ before transitioning along $e$ to $v'$. Denote this conditional holding time by variable $H(e)$. Here we assume that the holding time is dependent on both the current situation and the situation visited next. However, the conditional transition probabilities are independent of the holding times. Let $\pmb{\mathcal{H}}_\calT = \{\textbf{H}(v) | v \in S(\calT)\}$ where $\textbf{H}(v) = (H(e) | e = (v, v', l) \in E(\calT), v' \in \textmd{ch(v)})$ denotes the set of holding time variables for each edge emanating from situation $v \in S(\calT)$. Note that we assume here that all transitions in the event tree are associated with a holding time. For some temporal processes -- in particular, those including time-invariant covariates in their description -- might have some transitions for which a holding time is illogical. See \citet{shenvi2021non}, pp 87 -- 89, for a description of how these can be accommodated.

\begin{figure}
    \centering
    \includegraphics[scale = 0.3, trim = {0cm, 7cm, 0cm, 0cm}]{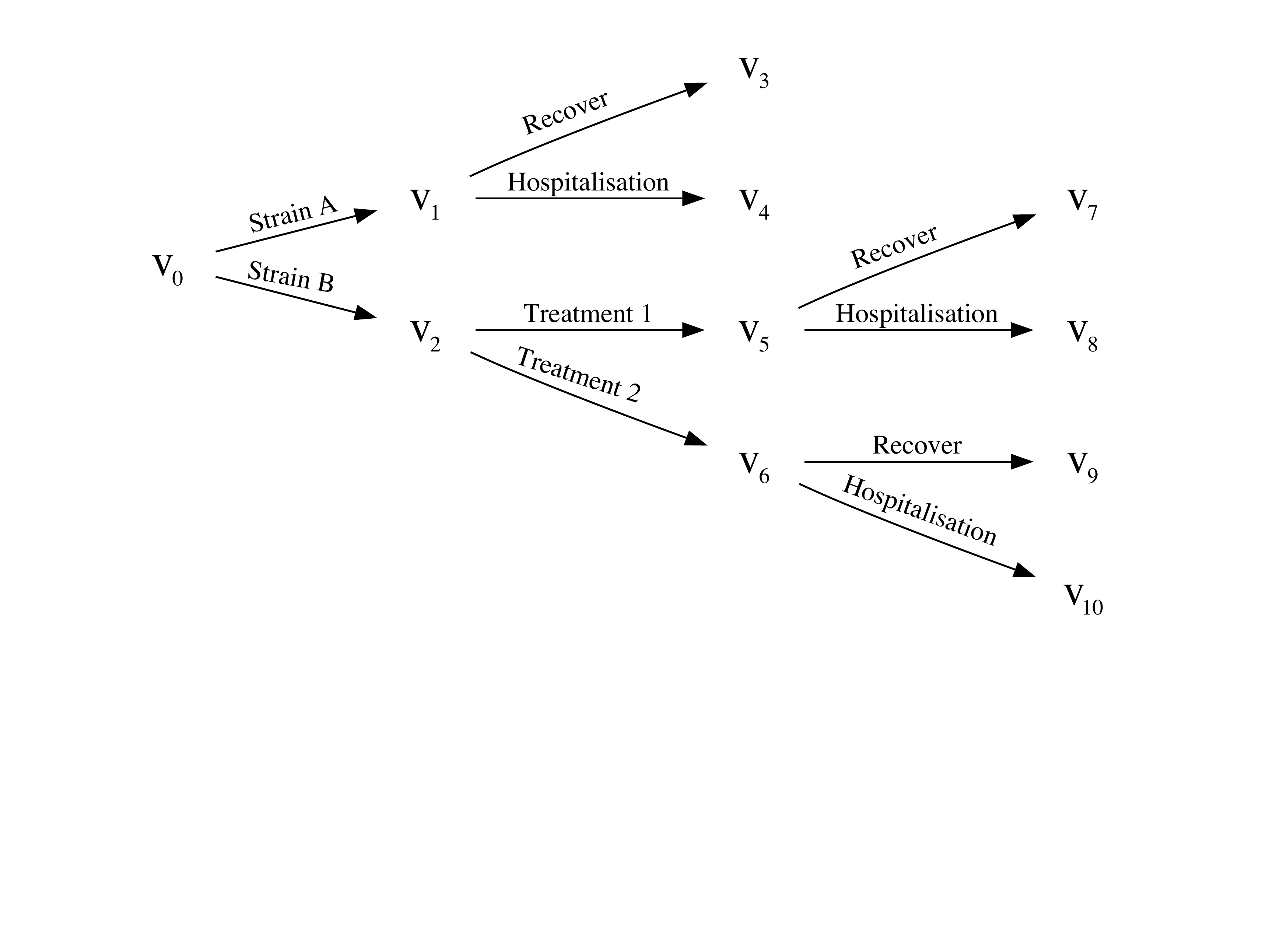}
    \caption{Event tree for the infection process in Example \ref{ex:infection}.}
    \label{fig:asymmetric_trees}
\end{figure}

\begin{example}[Infection example] \label{ex:infection}
Consider the infection example described earlier. Suppose we are studying hospitalisations occurring due to infection from one of two strains (strains A and B) of a circulating virus. Suppose that research showed that the available treatments are effective against an infection caused by strain B of the virus but not strain A. Therefore, individuals infected with strain A have no treatment options available whereas those infected with strain B have the options of treatment 1 and treatment 2. Thus, the treatment variable is structurally missing for individuals infected by strain A. The outcome of interest for this process is either recovery or hospitalisation. This process is structurally asymmetric and can be described by the event tree in Figure \ref{fig:asymmetric_trees}. Here, for situation $v_2 \in S(\calT)$ we have emanating edges $(v_2, v_5, \textmd{Treatment 1})$ and $(v_2, v_6, \textmd{Treatment 2})$, and its children are nodes $v_5$ and $v_6$. The random variables $H(v_2, v_5, \textmd{Treatment 1})$ and $H(v_2, v_6, \textmd{Treatment 2})$ describe the duration of the treatment after being infected by strain B for treatments 1 and 2 respectively. 

\end{example}

\begin{definition}[Stage] \label{def:stage}
In an event tree $\calT$, two situations $v$ and $v'$ are said to be in the same stage whenever
\begin{itemize}
\itemsep0em
\item $\pmb{\theta}_{v} = \pmb{\theta}_{v'}$ such that, for edges $e$ and $e'$ emanating from $v$ and $v'$ respectively with $\theta(e) = \theta(e')$, we require that $e = (v, \cdot, l)$ and $e' = (v', \cdot, l)$ for some edge label $l$;
\item Variables $H(e)$ and $H(e')$ for $e = (v, \cdot, l)$ and $e' = (v', \cdot, l)$ follow the same distribution. 
\end{itemize}
\end{definition}

Situations belonging to the same stage are given the same colour to represent the shared membership. An event tree $\calT$ whose situations are coloured according to their stage memberships is called a \textit{staged tree} and is denoted as $\calS$. The collection of stages $\mathbb{U}$ partitions the set of situations $S(\calT)$. It is common practice to suppress the colouring of trivial, i.e. singleton, stages to prevent visual cluttering.

Situations in the staged tree whose rooted subtrees are isomorphic have equivalent sets of edge labels, conditional transition parameters, and conditional holding time distributions\footnote{In a non-technical sense, this implies that $v$ and $v'$ have identical future evolutions.}. Situations whose rooted subtrees are isomorphic are said to belong to the same \textit{position}. Denote the collection of positions by $\mathbb{W}$. Observe that $\mathbb{W}$ creates a finer partition of $S(\calT)$. We can now define a temporal CEG as follows.

\begin{definition}[Temporal Chain Event Graph]
A temporal CEG $\calC = (V(\calC), E(\calC))$ is defined by the tuple $(\calS, \mathbb{W}, \pmb{\Phi}_\calS, \pmb{\mathcal{H}}_\calS)$ with the following properties:
\begin{itemize}
\itemsep0em
\item $V(\calC) = R(\mathbb{W}) \cup w_\infty$ where $R(\mathbb{W})$ is the set of situations representing each position set in $\mathbb{W}$ and $w_\infty$ is the \textit{sink} node. Additionally, nodes in $R(\mathbb{W})$ retain their stage colouring and for $w \in R(\mathbb{W})$, $\theta_\calC(w) = \theta_\calS(w)$ and $\textbf{H}_{\calC}(w) = \textbf{H}_{\calS}(w)$. 
\item Situations in $\calS$ belonging to the same position set in $\mathbb{W}$ are contracted into their representative node contained in $R(\mathbb{W})$. This node contraction merges multiple edges between two nodes into a single edge only if they share the same edge label.
\item Leaves of $\calS$ are contracted into sink node $w_\infty$.
\end{itemize}
\end{definition}

\begin{figure}
    \begin{subfigure}{0.45\textwidth}
    \centering
    \includegraphics[scale = 0.27, trim = {2cm, 6cm, 0cm, 0cm}]{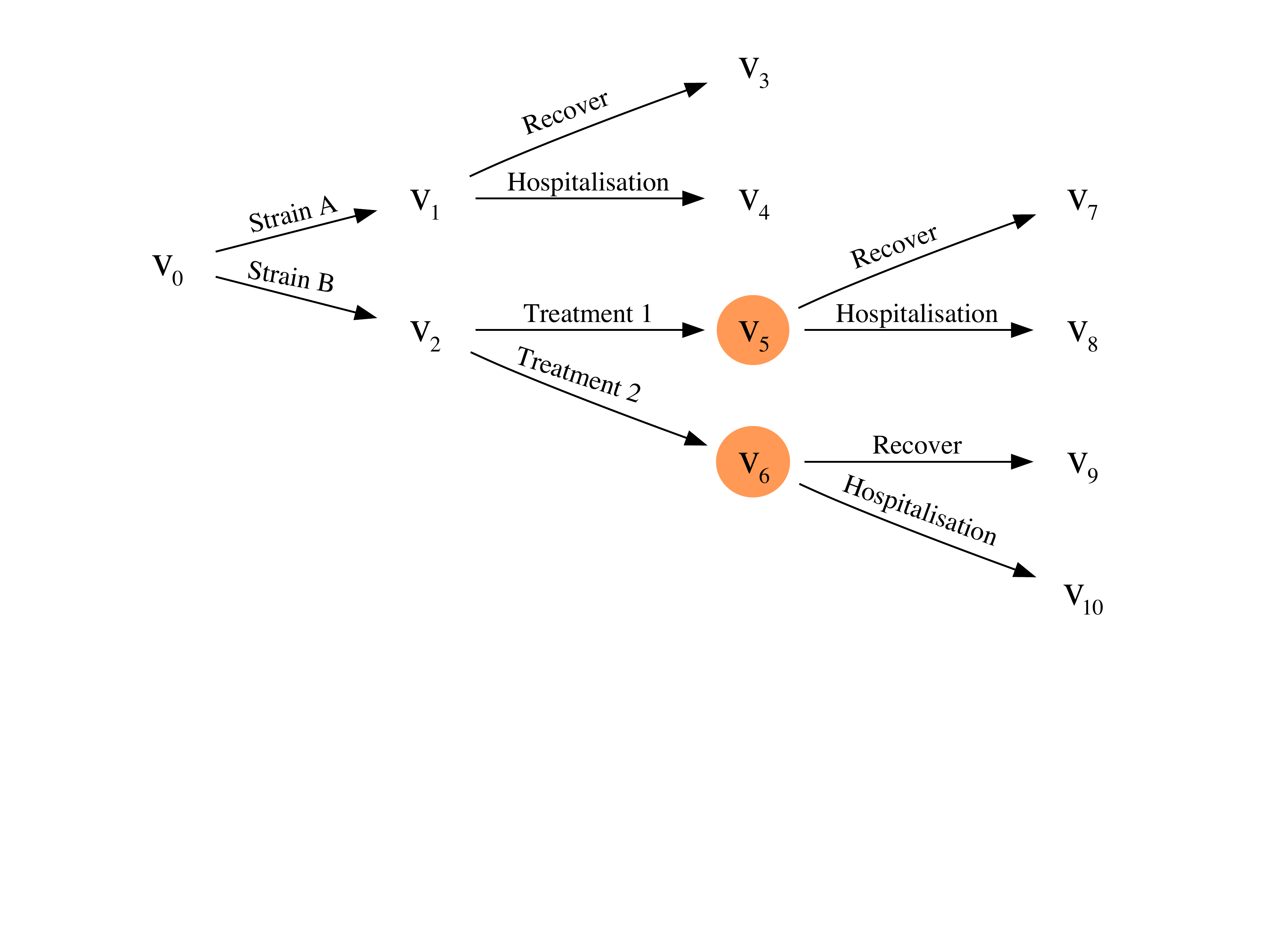}
    \caption{Staged tree}
    \end{subfigure}
    \begin{subfigure}{0.45\textwidth}
    \includegraphics[scale = 0.23, trim = {-4cm, 2.5cm, 0cm, 0cm}]{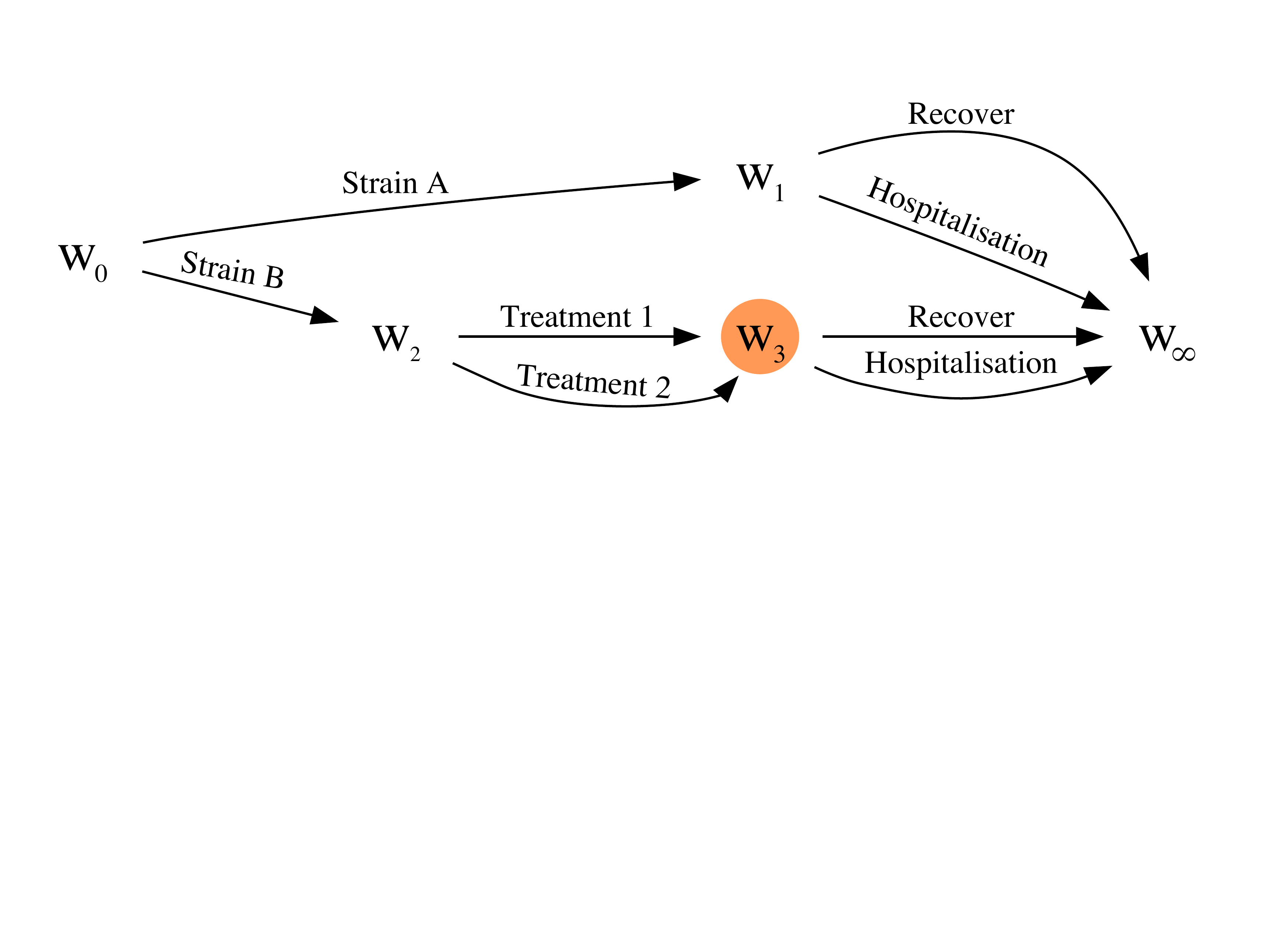}
    \caption{CEG}
    \end{subfigure}
    \caption{Staged tree and CEG for the infection process in Example \ref{ex:infection_staged}.}
    \label{fig:asymmetric_staged_ceg}
\end{figure}

\begin{example}[Infection example (continued)] \label{ex:infection_staged}
Suppose that the probability of recovery is independent of the treatment, given infection by strain B. This is a form of context-specific information which can be expressed as
\begin{align*}
    \textmd{Outcome} \ \indep \ \textmd{Treatment} \,|\, \textmd{Strain} = \textmd{Strain B}
\end{align*}
\noindent where $\indep$ stands for probabilistic independence and the vertical bar shows conditioning variables on the right. Suppose also that $H(v_1, v_3, \textmd{Recover})$, $H(v_5, v_7, \textmd{Recover})$ and $H(v_6, v_9, \textmd{Recover})$ follow the same distribution, and as do $H(v_1, v_4, \textmd{Hospitalisation})$, $H(v_5, v_8, \textmd{Hospitalisation})$ and $H(v_6, v_{10}, \textmd{Hospitalisation})$. The stage partition here is given by $\mathbb{U}$ which contains the following sets:
\begin{align*}
\{v_0\}, \{v_1\}, \{v_2\}, \{v_5, v_6\}.
\end{align*} 
\noindent Observe that $v_1$ is not in the same stage as $v_5$ and $v_6$ as although it satisfies the second condition given in Definition \ref{def:stage}, it does not satisfy the first. Figure \ref{fig:asymmetric_staged_ceg}(a) gives the staged tree for this process. In this example, the position partition $\mathbb{W}$ is equivalent to the stage partition $\mathbb{U}$. The leaves $v_3$, $v_4$, $v_7$, $v_8$, $v_9$ and $v_{10}$ are combined into a single sink node in the CEG as shown in Figure \ref{fig:asymmetric_staged_ceg}(b). 
\end{example}

\subsection{Separation of Likelihood}
\label{subsec:likelihood_separation}

We now demonstrate the conditions under which the parameters of the conditional transition and conditional holding time distributions can be learned independently. This separation of likelihood was first presented in \citet{barclay2015dynamic}.

Consider a temporal CEG $\calC$ with collection of stages $\mathbb{U} = \{u_1, u_2, \ldots, u_k\}$. Suppose that each stage $u_i$ has $k_i$ emanating edges (i.e. $\abs{\textmd{ch}(v_i)} = k_i$ for $v_i \in {u_i}$). Suppose we have a complete random sample of $n$ individuals. For each individual $1 \leq m \leq n$, let their data be given by the following sequence of tuples:
\begin{align*}
    \rho_m = ((e_{j_1k_1}, h_{j_1k_1}), (e_{j_2k_2}, h_{j_2k_2}), \ldots (e_{j_{l_m}k_{l_m}}, h_{j_{l_m}k_{l_m}})),
\end{align*}
\noindent where the first element of each tuple represents the edge traversed by the individual and the second element gives the holding time associated with that edge. Here we assume that for all individuals $j_1$ always corresponds to the root node of the CEG and that the final edge $e_{j_{l_m}k_{l_m}}$ for each individual $m$ ends in the sink node of the CEG. 

Denote the summary of the data associated with each stage $u_i$ in the sample by $\textbf{n}_i = (\textbf{n}_{i1}, \textbf{n}_{i2}, \ldots, \textbf{n}_{ik_i})$ and $\textbf{h}_i = (\textbf{h}_{i1}, \textbf{h}_{i2}, \ldots, \textbf{h}_{ik_i})$. Here, each $\textbf{n}_{ij}$ is a vector of ones of length $\abs{\textbf{n}_{ij}}$ where $\abs{\textbf{n}_{ij}}$ is the total number of individuals in the sample who traverse the $j$th edge of stage $u_i$. Correspondingly, $\textbf{h}_{ij}$ is a vector of the holding times for the $j$th edge of stage $u_i$ for each of the $\abs{\textbf{n}_{ij}}$ individuals in the sample who traverse this edge. 

The data from the $n$ individuals can now be summarised for the CEG as $\textbf{y} = \{\textbf{y}_1,\textbf{ y}_2, \ldots, \textbf{y}_k\}$ where $\textbf{y}_i = (\textbf{n}_i, \textbf{h}_i)$ corresponds to the data for stage set $u_i$, $i = 1, 2, \ldots, k$. 

Let the conditional transition parameters for stage $u_i$ be given by $\pmb{\theta}_i =\{\theta_{i1}, \theta_{i2}, \ldots, \theta_{ik_i}\}$ and let $\pmb{\Phi}_\calC = \{\pmb{\theta}_i| u_i \in \mathbb{U}\}$. Let the conditional holding time random variable for the $j$th edge emanating from stage $u_i$ be parametrised by $\pi_{ij}$. Then $\pmb{\pi}_i = \{\pi_{i1}, \pi_{i2}, \ldots, \pi_{ik_i}\}$ is the vector of holding time parameters for stage $u_i$. Let $\pmb{\Pi}_\calC = \{\pmb{\pi}_i| u_i \in \mathbb{U}\}$. The likelihood of the temporal CEG $\calC$ can be decomposed into a product of the likelihood of each stage as follows:
\begin{equation}
    p (\textbf{y}| \pmb{\Phi}_\calC, \pmb{\Pi}_\calC, \calC) = \prod_{i=1}^{k} p (\textbf{y}_i | \pmb{\theta}_i, \pmb{\pi}_i, \calC).
\end{equation}
We assume here that the conditional transition and conditional holding time parameters are \textit{a priori} mutually independent. This is analogous to the standard global parameter independence assumption in Bayesian networks \citep{spiegelhalter1990sequential} and vanilla CEGs \citep{freeman2011bayesian}. It follows under the separability of the likelihood above that they will also be independent \textit{a posteriori}. With this we can write
\begin{align}
    p (\textbf{y}_i | \pmb{\theta}_i, \pmb{\pi}_i, \calC) &= \prod_{j=1}^{k_i} p (\textbf{n}_{ij}, \textbf{h}_{ij} | {\theta}_{ij}, {\pi}_{ij}, \calC) \nonumber \\
    &= \prod_{j=1}^{k_i} p (\textbf{h}_{ij} | {\pi}_{ij}, \calC) p (\textbf{n}_{ij} | {\theta}_{ij}, \calC)  \nonumber \\
    &= \prod_{j=1}^{k_i} \prod_{l = 1}^{\abs{\textbf{n}_{ij}}} \Big \{ p ({h}_{ijl} | {\pi}_{ij}, \calC) \times p ({n}_{ijl} | {\theta}_{ij}, \calC) \Big \}.
\end{align}

Thus the likelihood of the model separates into the likelihoods of the conditional transition and conditional holding time parameters. This conveniently allows us to estimate the conditional transition and conditional holding time parameters independently. This holds irrespective of whether the conditional holding time variables are discrete or continuous. In the simulations in Section \ref{sec:experiments}, we demonstrate our methods for continuous conditional holding time variables. 

\subsection{CEG Model Selection}
\label{subsec:model_selection}

Model selection algorithms for temporal CEGs take as input the event tree $\calT$ of the process and output the staged tree $\calS$ for the process. A temporal CEG $\calC$ is uniquely and completely specified by its staged tree and the parameters over the staged tree $\pmb{\Phi}_{\calS}$ and $\pmb{\mathcal{H}}_\calS$ \citep{shenvi2020constructing}. Hence, the process of model selection in temporal CEGs is equivalent to identifying the collection of stages in its underlying event tree, which itself is identical to clustering the nodes of the event tree. Further, from Section \ref{subsec:likelihood_separation}, we can see that the process of clustering the nodes of the event tree can be split into two parts: 
\begin{enumerate}
    \itemsep0em
    \item \textbf{Identifying the situation clusters:} This refers to the first condition of a stage in Definition \ref{def:stage}. Here we aim to identify which sets of situations have equivalent conditional transition parameters. 
    \item \textbf{Identifying the edge clusters:} This refers to the second condition of a stage in Definition \ref{def:stage}. Here we aim to identify which sets of edges follow the same conditional holding time distribution. 
\end{enumerate}

The CEG model selection algorithms proposed in the literature are score-based and they fall under the two approaches of the greedy agglomerative hierarchical clustering (AHC) \citep{freeman2011bayesian, shenvi2019bayesian} and finding a globally optimal partition of the nodes with a dynamic programming approach \citep{cowell2014causal, silander2013dynamic}. Under both these approaches, the aim is to maximise a chosen score function. In the literature, this has generally taken the form of the log marginal likelihood score. The log marginal likelihood can be obtained analytically within the setting of conjugate priors for the conditional transition and conditional holding time distributions. Below we briefly outline the main steps involved in the AHC algorithm and the dynamic programming approach for identifying the situation clusters. These approaches can be similarly applied to identifying the edge clusters.  

The AHC algorithm is a local greedy search algorithm which aims to maximise the overall score by finding the next move that leads to a maximum increase in the score. It uses a bottom-up hierarchical clustering methodology beginning with the coarsest clustering treating each situation as a singleton cluster and successively merging pairs of clusters until the log marginal likelihood score cannot be improved further. The advantage of this approach is that it is fast when the number of situations is small or moderate. However, it is difficult to scale due to its cubic time complexity \citep{nielsen2016hierarchical}. Moreover, it does not scale robustly as it only searches a limited area of the model search space and can get stuck in a local maxima. For instance, a temporal CEG for a certain ordering of 4 binary variables -- each with the same set of edge labels -- has approximately $1.38 \times 10^9$ possible stagings but the AHC evaluates only 560 of them at most. In particular, once the AHC merges two situations into the same stage, it cannot undo this. Therefore, as the AHC algorithm is scaled, it tends to produce a large number of spurious clusters as we demonstrate later in Section \ref{subsec:situation_clusters}.

In contrast, the dynamic programming approach decomposes a larger problem (here, identifying situation clusters in the entire event tree) into smaller problems (identifying situation clusters within a given layer). If the event tree of the process being modelled has a symmetric structure -- also known as a stratified event tree in the literature -- then each layer corresponds to the set of situations that are at the same distance from the root of the stratified event tree \citep{cowell2014causal, silander2013dynamic}. If the process has an asymmetric structure, then the layers might be defined such that each layer contains situations belonging to the same underlying variable \citep{shenvi2018modelling}. To find the globally optimal partition using the dynamic programming approach, for each layer, we calculate the score for every partition of its situations and choose the partition with the highest score. This is a brute-force approach and is computationally very expensive. To see this, observe that the number of partitions to be evaluated for a layer with $k$ situations is given by the $k$th Bell number \citep{cowell2014causal} which grows exponential fast in $k$. 

\section{Mixture Models for CEG Model Selection}
\label{sec:model_selection}

In this section, we propose our novel model selection approach, based on mixture models, for temporal CEGs. This approach overcomes the limitation of assuming conjugate settings for the conditional transition and conditional holding time distributions, and is more amenable to robust scaling than the AHC algorithm described in Section \ref{subsec:model_selection}.  

\subsection{Mixture Models} 
\label{subsec:mixture_models}

We first briefly describe a finite mixture model. For an excellent exposition of finite mixture models see \citet{Frutthwirth-Schnatter}. Consider a population with $K$ subgroups where each subgroup $k$ is of relative proportion $\ell_k$, for $k = 1,2, \ldots, K$. Hence, $\sum_{k =1}^K \ell_k = 1$. Let $\pmb{\ell} = \{\ell_1, \ell_2, \ldots, \ell_K\}$. Suppose that the interest lies in modelling a random feature $Y$ such that $Y$ is heterogeneous across the subgroups but homogeneous within each subgroup. Hence, each subgroup $k$ can be associated with a parameter $\varphi_k$ for the distribution modelling $Y$; i.e. the distribution of $Y$ for subgroup $k$ is given by $p(Y= y \,|\, \varphi_k)$. Let $\pmb{\varphi} = \{\varphi_1, \varphi_2, \ldots, \varphi_K\}$.

Denote by $\textbf{y} = \{y_1, y_2, \ldots, y_n\}$ a random sample of feature $Y$ recorded from this population. Let an indicator variable $\pmb{z}_i = (z_i^1, z_i^2, \ldots, z_i^k)$ denote the subgroup occupied by an individual $i$ who is associated with the observation $y_i$. This gives us
\begin{equation*}
    z_i^k = \begin{cases}
    1, & \textmd{if $y_i$ comes from mixture component $k$,}\\
    0, & \textmd{otherwise.}
    \end{cases}
\end{equation*}
\noindent Assuming random sampling from the population, the probability that an individual belongs to subgroup $k$, for $1 \leq k \leq K$ is given by the Categorical distribution $Cat(\pmb{\ell})$. 




Typically, when we sample randomly from this population, we may not know which subgroup the individual belongs to. This could happen because of several reasons such as due to the way the data was collected or due to the subgroups being latent characteristics. The marginal density of $\textbf{y}$ here is given by the following mixture density
\begin{align}
    p(\textbf{y}) &= \prod_{i=1}^n p(y_i) \notag \\
    &= \prod_{i=1}^n \sum_{k=1}^K p(y_i, z_i^k) \notag \\
    &= \prod_{i=1}^n \sum_{k=1}^K p(z_i^k = 1 \,|\, \pmb{\ell})  \ p(y_i \,|\,z_i^k = 1, \varphi_k) \notag \\
    &= \prod_{i=1}^n \sum_{k=1}^K \ell_k p(y_i \,|\, \varphi_k).
    \label{eq:finite_mixture}
\end{align}

For finite mixture models with more than one mixture component (i.e. $K \geq 2$), the marginal likelihood $p(\textbf{y}\,|\, \mathscr{M})$ for some model $\mathscr{M}$ is not available in closed form and must be numerically approximated \citep{Frutthwirth-Schnatter}. 

We can evaluate the posterior probability of observation $y_i$, for an individual $i$, belonging to subgroup $k$ as follows
\begin{align}
    p(z_i^k = 1 | y_i) &= \dfrac{p(z^k_i = 1, y_i)}{p(y_i)} \notag \\
    &= \dfrac{p(z_i^k = 1 \,|\, \pmb{\ell}) \ p(y_i \,|\, z_i^k = 1, \varphi_k)}{ \sum_{j=1}^{K} p(z_i^j = 1 \,|\, \pmb{\ell}) \ p(y_i \,|\, z_i^j = 1, \varphi_j)} \notag \\
     &= \dfrac{\ell_k p(y_i \,|\, \varphi_k)}{ \sum_{j=1}^{K} \ell_j p(y_i \,|\, \varphi_j)}.
    \label{eq:posterior_allocation}
\end{align}
\noindent The above equation results in a soft clustering of the individuals. However, for most applications using CEGs, we are interested in a hard clustering. There are several ways of arriving at a hard clustering. In this paper, for posterior allocation of each individual $i$ to a single subgroup, we can choose the allocation as 
\begin{align}
    z^*_i = \mathop{\arg \max}\limits_{k \in \{1, 2, \ldots, K\}} p(z_i^k =1 \,|\, y_i).
    \label{eq:single_posterior_allocation}
\end{align}

\subsection{CEG Model Selection Approach Based on Mixture Models}
\label{subsec:model_selection_mixture_models}

We now describe how the CEG model selection problem can be cast as a mixture modelling problem. 

\subsubsection{Identifying the Situation Clusters}
\label{subsubsec:situation_clusters}

Consider an event tree $\calT$ with $n$ situations each with $m$ outgoing edges and the same set of edge labels. For situation $v_i \in S(\calT)$, let its associated data vector be given by ${\textbf{y}_i = (y_{i1}, y_{i2}, \ldots, y_{im})}$ where $y_{ij}$ represents the number of individuals in the random sample that arrive at situation $v_i$ and traverse its $j$th emanating edge, for $1 \leq i \leq n$ and $1 \leq j \leq m$. Here $\textbf{y} = \{\textbf{y}_1, \textbf{y}_2, \ldots, \textbf{y}_n\}$ is the data vector and $\pmb{\theta} = \{\pmb{\theta}_1, \pmb{\theta}_2, \ldots, \pmb{\theta}_n\}$ is the parameter vector where $\pmb{\theta}_i$ represents the conditional transition parameter vector for situation $v_i$. 

The model selection problem can be described as identifying the number and composition of the situation clusters in $\calT$. For a fixed number of situation clusters, this simplifies to fitting a standard finite mixture model as described by Equation \ref{eq:finite_mixture}. However, generally the number of situation clusters within a given event tree is unknown. To overcome this problem, we propose here an approach motivated by the AHC algorithm described in Section \ref{subsec:model_selection}. However, instead of a bottom-up approach like the AHC, we take a top-down approach\footnote{Note that a top-down approach with hierarchical clustering algorithms, known as divisive hierarchical clustering, is computationally very expensive with complexity typically being quartic or quintic \citep{roux2015comparative}.} as this generally results in a relatively conservative number of clusters. We start with fitting a mixture model with two clusters/components and then sequentially increase the number of components as long as there is an improvement in the log marginal likelihood score of the model. Recall that log marginal likelihood of a finite mixture model with two or more components is not available analytically. Instead, we estimate it using bridge sampling \citep{gronau2017tutorial}. A simplified pseudo-code of the proposed model selection algorithm is presented in Algorithm \ref{alg:situation_cluster}.

\begin{algorithm}[ht]
    \SetAlgoLined
    \Input{Data $\textbf{y}$, prior distribution for $\pmb{\theta}_i$ for $1 \leq i \leq n$, prior distribution for $\pmb{\ell}$.}
    \Output{Optimal number of situation clusters, collection of situation clusters.}
    Set \textit{allocation} $\leftarrow \emptyset$.\\
    Set \textit{parameters} $\leftarrow \emptyset$.\\
    Set \textit{score} $\leftarrow 0$.\\
    Set \textit{indicator} $\leftarrow 1$.\\
    Set \textit{k} $\leftarrow 2$.\\
	\While {\textit{indicator} $\neq$ 0} { 
	Fit the model as described by Equation \ref{eq:finite_mixture} with $k$ components.\\
	Set \textit{score}$_k$ as the log marginal likelihood of the fitted model using bridge sampling.\\
	\If {\textit{score}$_k$ $\geq$ \textit{score}} {
	\textit{score} $\leftarrow$ \textit{score}$_k$\\
	Set \textit{allocation} as the posterior allocation of each situation to one of the $k$ components as given by Equation \ref{eq:single_posterior_allocation}.\\
	Set \textit{parameters} as the mean posterior estimates of the parameters of each of the $k$ components.\\
	\textit{k} $\leftarrow$ \textit{k} + 1
	}
	\Else {
	\textit{indicator} $\leftarrow$ 0 }} 
	\KwRet{\textit{allocation}, \textit{parameters}} 
\caption{Mixture model selection algorithm for situation clusters}
\label{alg:situation_cluster}
\end{algorithm}

Whilst the above algorithm can easily handle several hundreds of situations for a fixed number of components, it will be significantly slowed down by fitting the mixture model for several potential number of components. As with the dynamic programming approach, the run time of the algorithm can be reduced by running it independently over suitably defined, mutually exclusive layers (see Section \ref{subsec:model_selection}).

In theory, the above algorithm is equally applicable for Binomial and Multinomial conditional transition distributions. However, fitting a Multinomial finite mixture in software such as Stan -- which we use for the experiments in Section \ref{sec:experiments} -- faces label switching problems among the components which can results in identifiability issues \citep{Frutthwirth-Schnatter, mena2015bayesian}. This is beyond the scope of this paper, and the subject of further research. Section \ref{sec:experiments} presents experiments for the Binomial case. We discuss possible approaches for circumventing the identifiability issues for the Multinomial finite mixture in Section \ref{sec:discussion}.

\subsubsection{Identifying the Edge Clusters}
\label{subsubsec:edge_clusters}

Identifying the edge clusters in an event tree requires a modification to the standard finite mixture modelling problem. Consider an event tree $\calT$ with $n$ edges which can all potentially be in the same edge cluster. For edge $e_i \in E(\calT)$, let $H(e_i)$ denote the conditional holding time random variable, for $1 \leq i \leq n$. Let $\textbf{y}_i = \{y_{i1}, y_{i2}, \ldots, y_{in_i}\}$ where $n_i$ indicates the number of individuals who traverse edge $e_i$ in our random sample and $y_{ij}$ represents the observed holding time for the $j$th individual traversing this edge, for $1 \leq i \leq n$ and $1 \leq j \leq n_i$. Let $\textbf{y} = \{\textbf{y}_1, \textbf{y}_2, \ldots, \textbf{y}_n\}$ be the data vector and $\pmb{\pi} = \{\pmb{\pi}_1, \pmb{\pi}_2, \ldots, \pmb{\pi}_n\}$ be the parameter vector where $\pmb{\pi}_i$ denotes the parameters associated with the conditional holding time distribution on edge $e_i$. Similar to the situation clusters in Section \ref{subsubsec:situation_clusters}, the model selection problem here can be described as identifying the number and composition of the edge clusters in $\calT$. However, in this case, we fit the non-standard mixture model given as 
\begin{align}
    p(\textbf{y})  &= \prod_{i=1}^n \sum_{k=1}^K \ell_k p(\textbf{y}_i \,|\, \pi_k) \notag \\
    &= \prod_{i=1}^n \sum_{k=1}^K \ell_k  \prod_{j=1}^{n_i} p(y_{ij} \,|\, \pi_k)
    \label{eq:finite_mixture_holding}
\end{align}
\noindent for a fixed number of components or clusters $K$. The mixture model given by Equation \ref{eq:finite_mixture_holding} is non-standard because contrary to conventional mixture models, it does not imply that each data observation (i.e. each observation of a holding time for any edge) independently comes from one of the mixture components. Observe here that our model implies that all the observed holding times in $\textbf{y}_i$ associated with edge $e_i$ necessarily belong to the same component. In other words, all the observations in $\textbf{y}_i$ are assumed to be drawn from the same distribution as they all correspond to holding times for the same edge. The model in Equation \ref{eq:finite_mixture_holding} can simply be viewed as a hierarchical model. 

The pseudo-code for this algorithm is identical to the pseudo-code in Algorithm \ref{alg:situation_cluster} with the exceptions that the adapted mixture model to be fit is given by Equation \ref{eq:finite_mixture_holding} and the posterior allocation is calculated as below
\begin{align}
    &z^*_i = \mathop{\arg \max}\limits_{k \in \{1, 2, \ldots, K\}} p(z_i^k =1 \,|\, \textbf{y}_i), \label{eq:edge_single_posterior_allocation} \\
    &\textmd{where} \quad p(z_i^k = 1 | \textbf{y}_i) = \dfrac{\ell_k \, \prod_{j=1}^{n_i} p(y_{ij} \,|\, \pi_k)}{ \sum_{m=1}^{K} \ell_m \prod_{j=1}^{n_i} p(y_{ij} \,|\, \pi_m)}. \notag
\end{align}

\section{Experiments} 
\label{sec:experiments}

In this section, we perform a series of computational experiments on simulated data to demonstrate the performance and properties of our proposed mixture modelling approach to model selection in CEGs. Similar to Section \ref{subsec:model_selection_mixture_models}, we consider the cases of identifying situation clusters and edge clusters separately. Throughout this section, we will use `stages/staging' to refer to the ground-truth clusters among the situations and edges, and `clusters/clustering' to refer to the clustering obtained by an algorithm. The experiments described in this section were run in R using the RStudio IDE on a 1.6 GHz MacBook Air with 8GB memory and were parallelised to run on 4 cores. The code for the experiments is provided as part of the supplementary materials.

\subsection{Situation Clusters} 
\label{subsec:situation_clusters}

Here, we compare the performance of our proposed methodology for identifying situation clusters, described in Section \ref{subsubsec:situation_clusters}, to that of the AHC algorithm. We shall only consider the Binomial case, i.e. where the conditional transition probabilities for the situations follow a Binomial distribution. We simulate 400 datasets for eight different scenarios (50 for each scenario) by setting the number of situations as 50, 200 or 450, and the number of generating stages as 2, 4 or 7 with the exception of the scenario with 50 situations and 7 stages. We do not consider the case of 50 situations and 7 stages as this results in some stages having very few data points which realistically makes it extremely difficult to identify the 7 stages correctly for any algorithm. While generating the datasets, the underlying Binomial success probabilities for the various stages are chosen to be distinct enough to minimise issues relating to identifiability \citep{Frutthwirth-Schnatter, mena2015bayesian}. Further, the number of situations belonging to the different stages is chosen at random for each simulation whilst ensuring that no stage has fewer than two situations. 

During each of the 50 simulations, for each of the eight datasets corresponding to the eight scenarios, we run the AHC algorithm, and the mixture modelling approach in Algorithm \ref{alg:situation_cluster} in Stan using the dataset. 

For the clustering obtained by AHC, we record the number of clusters, time taken (as clock-time in seconds) to run the algorithm and two measures of accuracy of the clustering compared to the ground-truth staging, namely, the normalised mutual information (NMI) score and the Rand index. The NMI score and the Rand index (see Appendix \ref{app:metrics} for more information) assesses the accuracy of the clustering labels compared to the ground-truth labels; for both of these, a score of 0 indicates poor clustering accuracy and 1 indicates perfect clustering. Recall here that the AHC algorithm begins by considering the coarsest partition where it treats each situation as a singleton cluster, and returns a hard clustering on the situations. 

To obtain the clustering from the mixture modelling approach (as described in Algorithm \ref{alg:situation_cluster}), we fit two or more Binomial mixture models in Stan. For fitting a Binomial mixture model in Stan, we run 4 chains, each with 1000 warmup iterations and 2000 post-warmup iterations. For the clustering obtained by the mixture modelling approach, we record the number of clusters, time taken (as clock-time in seconds), the NMI score, and the Rand index. Note here that unlike the AHC algorithm, the mixture model clustering begins with the finest partition, and returns a soft clustering on the situations through their posterior allocation probabilities. However, as described in Equation \ref{eq:single_posterior_allocation}, we choose a hard allocation of each situation to a single cluster. 

\begin{table}[ht]
\centering
\begin{tabular}{cccccc}
  \hline
 \# Situations & \# Stages & Time Taken & \# Clusters & NMI Score & Rand Index\\  
  \hline
50 & 2 & 3.60 & 4.52 & 0.59 & 0.61 \\ 
50 & 4 & 3.11 & 6.20 & 0.67 & 0.78 \\ 
200 & 2 & 101.07 & 7.80 & 0.49 & 0.49 \\ 
200 & 4 & 99.70 & 10.62 & 0.57 & 0.72 \\ 
200 & 7 & 97.00 & 14.80 & 0.72 & 0.84 \\ 
450 & 2 & 1144.17 & 10.52 & 0.45 & 0.45  \\ 
450 & 4 & 1134.43 & 14.64 & 0.55 & 0.71 \\ 
450 & 7 & 1121.28 & 20.08 & 0.68 & 0.82 \\ 
   \hline
\end{tabular}
\caption{Summary of results for clustering the situations, with underlying Binomial conditional transition distributions, using the AHC algorithm. }
\label{tab:AHC}
\end{table}

\begin{table}[ht]
\centering
\begin{tabular}{ccccccc}
  \hline
 \# Situations & \# Stages & Time Taken & \# Clusters & NMI Score & Rand Index \\ 
  \hline
50 & 2 & 70.29 & 2.00 & 0.96 & 0.98 \\ 
50 & 4 & 159.13 & 3.50 & 0.78 & 0.84 \\ 
200 & 2 & 227.23 & 2.08 & 0.98 & 0.99 \\ 
200 & 4 & 780.36 & 3.90 & 0.75 & 0.84 \\ 
200 & 7 & 1235.65 & 5.82 & 0.87 & 0.93 \\ 
450 & 2 & 450.37 & 2.08 & 0.98 & 0.98 \\ 
450 & 4 & 3036.13 & 4.52 & 0.74 & 0.84 \\ 
450 & 7 & 4039.94 & 6.20 & 0.85 & 0.92 \\ 
   \hline
\end{tabular}
\caption{Summary of results for clustering the situations, with underlying Binomial conditional transition distributions, using the mixture modelling approach. }
\label{tab:Stan}
\end{table}

The summary of the results is presented in Table \ref{tab:AHC} for the AHC algorithm and in Table \ref{tab:Stan} for the mixture model clustering. Each scenario is defined by the number of situations (reported as \# Situations) and the number of underlying stages (reported as \# Stages). The time taken, number of clusters (reported as \# Clusters), NMI score and Rand index are averaged over the 50 simulations for each scenario. In most cases, the mixture model clustering takes a considerably longer time than the AHC, with the exception of the scenario with 450 situations and 2 underlying stages. This occurs due to the difference in the AHC's top-down approach and the mixture model clustering's bottom-up approach. However, the mixture model clustering consistently performs better than the AHC in terms of the clustering accuracy metrics. The summary of the convergence results for the simulations is presented in Appendix \ref{app:convergence}.



\subsection{Edge Clusters} 
\label{subsec:edge_clusters}

For the edge clusters, we analyse the performance of our mixture modelling approach in the case where we do not have conjugacy. In this case, the AHC algorithm as described in Section \ref{subsec:model_selection} is not applicable and hence, we cannot use it for comparative purposes. Here, the conditional holding time data for each edge is assumed to come from a Weibull distribution with known scale parameter and unknown shape parameter. Recall that the Weibull distribution only enjoys a conjugate prior for the scale parameter when the shape parameter is known and the scale parameter is unknown. Similar to the setting in Section \ref{subsec:situation_clusters}, we simulate 400 datasets for eight different scenarios (50 for each scenario) by setting the number of edges as 50, 200 or 450, and the number of generating stages to 2, 4 or 7 with the exception of the scenario with 50 edges and 7 stages. We set the scale parameters for all stages to be 50 and set the underlying shape parameters for the different stages to be distinct enough to minimise identifiability issues. Each row in the dataset is a vector of 30 conditional holding time observations for the edge it represents. The number of edges belonging to the different stages is chosen at random for each simulation whilst ensuring that no stage has fewer than five edges. 

For each clustering obtained by the mixture modelling approach (as described in Section \ref{subsubsec:edge_clusters}), we fit a Weibull mixture model in Stan with known scale parameter and estimate the unknown shape parameter. We fit the model using 4 chains, each with 1000 warmup iterations and 2000 post-warmup iterations. We record the number of clusters, time taken (as clock-time in seconds), the NMI score, and the Rand index for the clustering obtained through the approach. We enforce a hard clustering as described in Equation \ref{eq:edge_single_posterior_allocation}. The summarised results averaged over the 50 simulations for each of the eight scenarios are presented in Table \ref{tab:holding_stan}. This approach has very good performance as evidenced by average values of the number of clusters (reported as \# Clusters), NMI score and Rand index for all eight scenarios, in particular when the underlying stages are 2 or 4. As for the situation clusters, the summary of the convergence results is presented in Appendix \ref{app:convergence}.

\begin{table}[ht]
\centering
\begin{tabular}{cccccc}
  \hline
 \# Edges & \# Stages & Time Taken & \# Clusters & NMI Score & Rand Index \\ 
  \hline
50 & 2 & 133.01 & 2.00 & 1.00 & 1.00 \\ 
  50 & 4 & 353.11 & 4.00 & 0.99 & 1.00 \\ 
  200 & 2 & 676.19 & 2.00 & 1.00 & 1.00 \\ 
  200 & 4 & 2034.83 & 4.04 & 0.99 & 1.00 \\ 
  200 & 7 & 4623.07 & 6.28 & 0.88 & 0.93 \\ 
  450 & 2 & 1879.68 & 2.00 & 1.00 & 1.00 \\ 
  450 & 4 & 9260.44 & 4.02 & 0.99 & 1.00 \\ 
  450 & 7 & 11961.49 & 5.32 & 0.81 & 0.87 \\ 
   \hline
   \end{tabular}
   \caption{Summary of results for clustering the edges, with underlying Weibull conditional holding time distributions with known scale parameters and unknown shape parameters, using the mixture modelling approach. }
   \label{tab:holding_stan}
\end{table}

\section{Discussion} 
\label{sec:discussion}

In this paper we have shown that by viewing model selection for CEGs as a clustering problem, we can use a mixture modelling approach for model selection in CEGs. We demonstrated that this approach is very promising when the conditional holding time distributions do not have conjugate priors and also for robustly scaling to a larger number of situations (or equivalently, edges) as compared to the AHC algorithm even under the assumption of conjugacy. 

This work opens up several avenues for future work; most excitingly for new applications of CEGs for processes with arbitrary holding time distributions and/or a large number of nodes in its event trees. Further, the soft clustering provided naturally by the mixture modelling approach can be used with a Bayesian model averaging setting such as in \citet{strong2022bayesian} for robust explanatory analyses using a set of top-scoring models rather than just the maximum \textit{a posteriori} model.   

There are challenges that will require further study. The conditional probability distribution for a situation with three or more emanating edges follows a Multinomial distribution. Fitting a Multinomial mixture model in Stan faces identifiability issues. \citet{identifiability_guide} recommends identifying degenerate Bayesian mixture models by either using non-exchangeable priors or enforcing an ordering on the parameters. In the Binomial case, we used the latter approach on the probability of success parameter of the Binomial distribution. However, for the Multinomial case, enforcing an ordering is not sufficient as for a Multinomial with $k$ categories has degree of freedom $k-1$ and it is not straightforward how to enforce an ordering on all $k-1$ categories at once. There are two possible approaches that we could consider for further study. One being that a Multinomial distribution can be written as a series of consecutive Binomial distributions and the other that for a specific application, non-exchangeable priors could be used. 


Secondly, our current approach scales well in the number of situations or edges in the tree but estimating number of components using current method is not easily scalable as the number of underlying stages increases. Within a specific application, in order to minimise the computational load, it is advisable to elicit a suitable range for the number of components prior to commencing the model selection process. 

Finally, in this paper, we only considered the case where one of the two parameters of the Weibull distribution was unknown. This is due to within chain and between chain label switching observed in our simulations when we assumed both parameters to be unknown. However, in a specific application, this can be easily ameliorated by using techniques such as non-exchangeable priors, using parameter constraints (such as enforcing an ordering), choosing a smaller window of values to consider for the number of components if it is large and performing post-hoc analysis to correct label switching issues after the model has been fit (see e.g. \citet{cassiday2021comparison, almond2014comparison}.


Despite these challenges that require further study, the approach that we propose in this paper vastly extends the applicability of CEGs and will open up a range of opportunities.

\section*{Appendix}
\renewcommand{\thesubsection}{\Alph{subsection}}

\subsection{Clustering Accuracy Metrics} \label{app:metrics}

The normalised mutual information (NMI) score and the Rand index are two popular metrics for comparing the accuracy of a clustering algorithm. Let $GT$ denote the ground-truth or generating cluster labels of the data points and $Pred$ denote their corresponding predicted cluster labels. The NMI is a normalisation of the mutual information score and it is obtained as follows:
\begin{align*}
    NMI(Pred, GT) = \frac{I(Pred, GT)}{\sqrt{H(Pred) H(GT)}},
\end{align*}
\noindent where $I(Pred, GT)$ denotes the mutual information between the two labellings and $H(Pred)$ denotes the entropy of $Pred$. Here, a score of 0 indicates no mutual information whereas a score of 1 indicates perfect correlation. 

The Rand index measures the percentage of correct decisions made by the clustering algorithm and is given as
\begin{align*}
    RI(Pred, GT) = \frac{TP + TN}{TP + FP + TN + FN},
\end{align*}
\noindent where $TP, TN, FP$ and $FN$ are the true positives, true negatives, false positives and false negatives respectively in $Pred$ compared to $GT$. The range of the Rand index is $[0,1]$ with a higher value indicating a better clustering accuracy. 

\subsection{Convergence Results for the Experiments} \label{app:convergence}

\subsubsection{Situation Clusters}
\begin{table}[ht]
\centering
\begin{tabular}{cccc}
  \hline
\# Situations & \# Stages & Prop Converging Lvl1 & Prop Converging Lvl2 \\ 
  \hline
50 & 2 & 0.96 & 0.96 \\ 
50 & 4 & 0.83 & 0.95 \\ 
200 & 2 & 0.90 & 0.90 \\ 
200 & 4 & 0.43 & 0.58 \\ 
200 & 7 & 0.68 & 0.75 \\ 
450 & 2 & 0.92 & 0.94 \\ 
450 & 4 & 0.36 & 0.53 \\ 
450 & 7 & 0.50 & 0.63 \\ 
   \hline
\end{tabular}
\caption{Summary of the convergence results for clustering the situations using the mixture modelling approach.}
\label{tab:conv1}
\end{table}

Unlike the AHC algorithm which uses closed form equations to estimate the parameters of interest, the mixture model clustering implemented in Stan estimates the parameters of interest using a No-U-Turn Sampler (NUTS) \citep{carpenter2017stan}. Hence, as a diagnostic check we analyse whether the parameters relating to the Binomial distribution for each component converge. The computation is said to converge when the split-$\hat{R} < 1.01$ as recommended by \citet{vehtari2021rank}. This is a much tighter bound compared to the original recommended bound of 1.10 \citep{gelman1992inference}. For comparative purposes, we also check convergence under the $1.10$ threshold. Table \ref{tab:conv1} shows the proportion of Binomial parameters that converged at the threshold of 1.01 (reported as Prop Converging Lvl1) and 1.10 (reported as Prop Converging Lvl2) for each scenario averaged over the 50 simulations. Over two-thirds of the parameters converged under both thresholds for each scenario except for the scenarios of 200 situations \& 4 stages and 450 situations \& 4 or 7 stages. 

\subsubsection{Edge Clusters}

\begin{table}[ht]
\centering
\begin{tabular}{cccc}
  \hline
\# Edges & \# Stages & Prop Converging Lvl1 & Prop Converging Lvl2 \\ 
  \hline
50 & 2 & 1.00 & 1.00 \\ 
50 & 4 & 1.00 & 1.00 \\ 
200 & 2 & 1.00 & 1.00 \\ 
200 & 4 & 0.98 & 1.00 \\ 
200 & 7 & 0.56 & 0.67 \\ 
450 & 2 & 1.00 & 1.00 \\ 
450 & 4 & 0.99 & 1.00 \\ 
450 & 7 & 0.26 & 0.34 \\ 
   \hline
\end{tabular}
\caption{Summary of the convergence results for clustering the edges using the mixture modelling approach.}
\label{tab:conv2}
\end{table}

We analyse the convergence properties of the mixture modelling approach to clustering the edges under the thresholds of 1.01 and 1.10 for the split-$\hat{R}$. This is summarised in Table \ref{tab:conv2} with the threshold of 1.01 reported as Prop Converging Lvl1 and that of 1.10 reported as Prop Converging Lvl2. Almost all 50 simulations converged for each scenario, with the exception of 200 edges \& 7 stages and 450 edges \& 7 stages. 

The convergence results when we have 7 stages are not as good as for fewer stages. Recall that we compared models using their log marginal likelihoods which were approximated using bridge sampling. This is equivalent to using the Bayes Factor \citep{kass1995bayes} where all the models are \textit{a priori} equally likely. The Bayes Factor, whilst an extremely common approach to model comparison, has several drawbacks. It is very sensitive to priors and difficult to approximate accurately \citep{bridgesampling, schad2022workflow, oelrich2020bayesian}. Therefore, the approximated Bayes Factor is not always suitable for comparing models especially when the approximation is carried out in a black-box manner. Further, observe that when the number of estimated components is two, we only estimate two log marginal likelihoods and make one Bayes Factor comparison. However, when the estimated number of components is 7, we have 6 log marginal likelihood estimations and 5 Bayes Factor comparisons; thereby increasing the possibility of errors caused due to the use of Bayes Factors. In practice, we recommend careful checks of the Stan and bridgesampling outputs, and the use of post-hoc analysis if necessary; see Section \ref{sec:discussion}.


\printbibliography[title= References]

@article{freeman2011bayesian,
  title={Bayesian {MAP} model selection of chain event graphs},
  author={Freeman, Guy and Smith, Jim Q},
  journal={Journal of {M}ultivariate {A}nalysis},
  volume={102},
  number={7},
  pages={1152--1165},
  year={2011}
}

@article{roux2015comparative,
  title={A comparative study of divisive hierarchical clustering algorithms},
  author={Roux, Maurice},
  journal={ar{X}iv:1506.08977},
  year={2015}
}

@inproceedings{shenvi2018modelling,
  title={Modelling with non-stratified chain event graphs},
  author={Shenvi, Aditi and Smith, Jim Q and Walton, Robert and Eldridge, Sandra},
  booktitle={International {C}onference on {B}ayesian {S}tatistics in {A}ction},
  pages={155--163},
  year={2018},
  organization={Springer}
}

@inproceedings{shenvi2020constructing,
    title={Constructing a chain event graph from a staged tree},
    author={Shenvi, Aditi and Smith, Jim Q},
    booktitle={Proceedings of the {T}enth {I}nternational {C}onference on {P}robabilistic {G}raphical {M}odels},
    year={2020},
  organization={PMLR}
}

@phdthesis{shenvi2021non, 
    title={Non-Stratified Chain Event Graphs: Dynamic Variants, Inference and Applications}, 
    author={Shenvi, Aditi}, 
    school={The {U}niversity of {W}arwick},
    year={2021}
}

@techreport{minka2003bayesian,
  title={Bayesian inference, entropy, and the multinomial distribution},
  author={Minka, Thomas P},
  institution={Microsoft {R}esearch},
  year={2003}
}

@inproceedings{strong2022bayesian,
  title={Bayesian Model Averaging of Chain Event Graphs for Robust Explanatory Modelling},
  author={Strong, Peter and Smith, Jim Q},
  booktitle={Proceedings of the {E}leventh {I}nternational {C}onference on {P}robabilistic {G}raphical {M}odels},
  year={2022},
  organization={PMLR}
}

@article{barclay2015dynamic,
  title={The dynamic chain event graph},
  author={Barclay, Lorna M and Collazo, Rodrigo A and Smith, Jim Q and Thwaites, Peter A and Nicholson, Ann E},
  journal={Electronic {J}ournal of {S}tatistics},
  volume={9},
  number={2},
  pages={2130--2169},
  year={2015}
}

@inproceedings{jabbari2018instance,
  title={Instance-specific {B}ayesian network structure learning},
  author={Jabbari, Fattaneh and Visweswaran, Shyam and Cooper, Gregory F},
  booktitle={Proceedings of the {N}inth {I}nternational {C}onference on {P}robabilistic {G}raphical {M}odels},
  pages={169--180},
  year={2018},
  organization={PMLR}
}

@article{poole2003exploiting,
  title={Exploiting contextual independence in probabilistic inference},
  author={Poole, David and Zhang, Nevin L},
  journal={Journal of {A}rtificial {I}ntelligence {R}esearch},
  volume={18},
  pages={263--313},
  year={2003}
}

@inproceedings{zhang1999role,
  title={On the role of context-specific independence in probabilistic inference},
  author={Zhang, Nevin L and Poole, David},
  booktitle={Proceedings of the 16th {I}nternational {J}oint {C}onference on {A}rtificial {I}ntelligence},
  volume = {2},
  pages={1288--1293},
  year={1999}
}

@inproceedings{boutilier1996context,
  title={Context-specific independence in {B}ayesian networks},
  author={Boutilier, Craig and Friedman, Nir and Goldszmidt, Moises and Koller, Daphne},
  booktitle={Proceedings of the {T}welfth {I}nternational {C}onference on {U}ncertainty in {A}rtificial {I}ntelligence},
  pages={115--123},
  year={1996}
}

@inproceedings{silander2013dynamic,
  title={A dynamic programming algorithm for learning chain event graphs},
  author={Silander, Tomi and Leong, Tze-Yun},
  booktitle={International {C}onference on {D}iscovery {S}cience},
  pages={201--216},
  year={2013},
  organization={Springer}
}

@article{cowell2014causal,
  title={Causal discovery through {MAP} selection of stratified chain event graphs},
  author={Cowell, Robert G and Smith, Jim Q},
  journal={Electronic {J}ournal of {S}tatistics},
  volume={8},
  number={1},
  pages={965--997},
  year={2014}
}

@article{smith2008conditional,
  title={Conditional independence and chain event graphs},
  author={Smith, Jim Q and Anderson, Paul E},
  journal={Artificial {I}ntelligence},
  volume={172},
  number={1},
  pages={42--68},
  year={2008},
  publisher={Elsevier}
}

@book{Frutthwirth-Schnatter,
    author = {Fr\"{u}hwirth-Schnatter, Sylvia},
    year  = {2006},
    publisher = {Springer},
    title = {Finite mixture and {M}arkov switching models}
}

@article{mena2015bayesian,
  title={On the {B}ayesian mixture model and identifiability},
  author={Mena, Rams{\'e}s H and Walker, Stephen G},
  journal={Journal of {C}omputational and {G}raphical {S}tatistics},
  volume={24},
  number={4},
  pages={1155--1169},
  year={2015},
  publisher={Taylor \& {F}rancis}
}

@article{carpenter2017stan,
  title={Stan: A probabilistic programming language},
  author={Carpenter, Bob and Gelman, Andrew and Hoffman, Matthew D and Lee, Daniel and Goodrich, Ben and Betancourt, Michael and Brubaker, Marcus and Guo, Jiqiang and Li, Peter and Riddell, Allen},
  journal={Journal of statistical software},
  volume={76},
  number={1},
  pages={1--32},
  year={2017}
}

@misc{identifiability_guide,
    title = {Identifying Bayesian Mixture Models},
    author = {Betancourt, Michael},
    url = {https://betanalpha.github.io/assets/case_studies/identifying_mixture_models.html},
    year = {2017}
}

@article{oelrich2020bayesian,
  title={When are Bayesian model probabilities overconfident?},
  author={Oelrich, Oscar and Ding, Shutong and Magnusson, M{\aa}ns and Vehtari, Aki and Villani, Mattias},
  journal={arXiv preprint arXiv:2003.04026},
  year={2020}
}

@article{schad2022workflow,
  title={Workflow techniques for the robust use of bayes factors.},
  author={Schad, Daniel J and Nicenboim, Bruno and B{\"u}rkner, Paul-Christian and Betancourt, Michael and Vasishth, Shravan},
  journal={Psychological Methods},
  year={2022},
  publisher={American Psychological Association}
}

@article{kass1995bayes,
  title={Bayes factors},
  author={Kass, Robert E and Raftery, Adrian E},
  journal={Journal of the american statistical association},
  volume={90},
  number={430},
  pages={773--795},
  year={1995},
  publisher={Taylor \& Francis}
}

@inproceedings{almond2014comparison,
  title={A comparison of two {MCMC} algorithms for hierarchical mixture models},
  author={Almond, Russell G},
  booktitle={Proceedings of the {E}leventh {UAI} {C}onference on {B}ayesian {M}odeling {A}pplications {W}orkshop},
  volume = {1218},
  pages={1--19},
  year={2014}
}

@article{cassiday2021comparison,
  title={A comparison of label switching algorithms in the context of growth mixture models},
  author={Cassiday, Kristina R and Cho, Youngmi and Harring, Jeffrey R},
  journal={Educational and Psychological Measurement},
  volume={81},
  number={4},
  pages={668--697},
  year={2021},
  publisher={Sage Publications Sage CA: Los Angeles, CA}
}

@article{vehtari2021rank,
  title={Rank-Normalization, Folding, and Localization: {A}n Improved {$\hat{R}$} for Assessing Convergence of {MCMC}},
  author={Vehtari, Aki and Gelman, Andrew and Simpson, Daniel and Carpenter, Bob and B{\"u}rkner, Paul-Christian},
  journal={Bayesian Analysis},
  volume={1},
  number={1},
  pages={1--28},
  year={2021},
  publisher={International Society for Bayesian Analysis}
}

@article{gelman1992inference,
  title={Inference from iterative simulation using multiple sequences},
  author={Gelman, Andrew and Rubin, Donald B},
  journal={Statistical science},
  volume={7},
  number={4},
  pages={457--472},
  year={1992},
  publisher={Institute of Mathematical Statistics}
}

@article{collazo2018n,
  title={An {N} time-slice dynamic chain event graph},
  author={Collazo, Rodrigo A and Smith, Jim Q},
  journal={arXiv:1808.05726},
  year={2018}
}

@book{shafer1996art,
  title={The art of causal conjecture},
  author={Shafer, Glenn},
  year={1996},
  publisher={{MIT} {P}ress}
}

@article{shenvi2019bayesian,
  title={A {B}ayesian dynamic graphical model for recurrent events in public health},
  author={Shenvi, Aditi and Smith, Jim Q},
  year={2019},
  journal = {ar{X}iv:1811.08872}
}

@book{collazo2018chain,
  title={Chain event graphs},
  author={Collazo, Rodrigo A and G{\"o}rgen, Christiane and Smith, Jim Q},
  year={2018},
  publisher={{CRC} {P}ress}
}

@incollection{nielsen2016hierarchical,
  title={Hierarchical clustering},
  author={Nielsen, Frank},
  booktitle={Introduction to HPC with MPI for Data Science},
  pages={195--211},
  year={2016},
  publisher={Springer}
}

@article{spiegelhalter1990sequential,
  title={Sequential updating of conditional probabilities on directed graphical structures},
  author={Spiegelhalter, David J and Lauritzen, Steffen L},
  journal={Networks},
  volume={20},
  number={5},
  pages={579--605},
  year={1990},
  publisher={Wiley {O}nline {L}ibrary}
}

@Article{bridgesampling,
    title = {{bridgesampling}: An {R} Package for Estimating Normalizing Constants},
    author = {Quentin F. Gronau and Henrik Singmann and Eric-Jan Wagenmakers},
    journal = {Journal of Statistical Software},
    year = {2020},
    volume = {92},
    number = {10},
    pages = {1--29},
    doi = {10.18637/jss.v092.i10},
  }

@article{gronau2017tutorial,
  title={A tutorial on bridge sampling},
  author={Gronau, Quentin F and Sarafoglou, Alexandra and Matzke, Dora and Ly, Alexander and Boehm, Udo and Marsman, Maarten and Leslie, David S and Forster, Jonathan J and Wagenmakers, Eric-Jan and Steingroever, Helen},
  journal={Journal of mathematical psychology},
  volume={81},
  pages={80--97},
  year={2017},
  publisher={Elsevier}
}

\end{document}